\documentclass[11pt,a4paper]{article}
\usepackage{amsmath,amssymb}
\usepackage{epsfig,graphicx}
\usepackage{cite}

\begin{document}

\title{
\begin{flushright}
{\normalsize Yaroslavl State University\\
             Preprint YARU-HE-06/03\\
             hep-ph/0606262} \\[10mm]
\end{flushright}
PLASMA INDUCED NEUTRINO RADIATIVE DECAY\\ INSTEAD OF NEUTRINO SPIN LIGHT} 

\author{A.~V.~Kuznetsov$^a$\footnote{{\bf e-mail}: avkuzn@uniyar.ac.ru},
N.~V.~Mikheev$^{a}$\footnote{{\bf e-mail}: mikheev@uniyar.ac.ru}
\\
$^a$ \small{\em Yaroslavl State (P.G.~Demidov) University} \\
\small{\em Sovietskaya 14, 150000 Yaroslavl, Russian Federation}
}

\date{}

\maketitle

\begin{abstract}
The conversion $\nu_L \to \nu_R \gamma^*$ of a neutrino with a magnetic 
moment is considered, caused by the additional Wolfenstein energy acquired by 
a left-handed neutrino in medium, 
with an accurate taking account of the photon $\gamma^*$ dispersion in medium. 
It is shown that the threshold arises in the process, caused by the photon 
(plasmon) effective mass. This threshold leaves no room for the so-called 
``neutrino spin light'' in the most of astrophysical situations. \\


\end{abstract}
 
\vfill

\begin{center}
 {\it Submitted to Modern Physics Letters A} 
\end{center}


\newpage

The most important event in neutrino physics of the last decades was 
the solving of the Solar neutrino problem, made in the unique experiment 
on the heavy-water detector at the Sudbury Neutrino Observatory~\cite{SNO:2001}. 
This experiment, together with the atmospheric and the reactor neutrino 
experiments~\cite{SK:1998,KamLAND:2003}, has confirmed the key idea by 
B. Pontecorvo on neutrino oscillations~\cite{Pontecorvo:1957}. 
The existence of non-zero neutrino mass and lepton mixing is thereby established. 
The Sun appeared in this case as a natural laboratory for investigations 
of neutrino properties. 

There exists a number of natural laboratories, the supernova explosions, where 
gigantic neutrino fluxes define in fact the process energetics. 
It means that microscopic neutrino characteristics, such as the neutrino 
magnetic moment, the neutrino dispersion in an active medium, etc., would have 
a critical impact on macroscopic properties of these astrophysical events.

This is the reason for a growing interest to neutrino physics in an external 
active medium. In an astrophysical environment, the main medium influence 
on neutrino properties is defined by the additional Wolfenstein 
energy $W$ acquired by a left-handed neutrino~\cite{Wolfenstein:1978}. 
The general expression for this additional energy of 
a left-handed neutrino with the flavor $i = e, \mu, \tau$ 
is~\cite{Notzold:1988,Pal:1989,Nieves:1989}
\begin{eqnarray}
&&W_i = \sqrt{2} \, G_{\rm F} \left[
\left(\delta_{ie} - \frac{1}{2} + 2 \, \sin^2 \theta_{\rm W}\right) 
\left(N_e - \bar N_e \right) \right.
\nonumber\\
&&+ \left. \left(\frac{1}{2} - 2 \, \sin^2 \theta_{\rm W}\right) 
\left(N_p - \bar N_p \right) 
- \frac{1}{2} \left(N_n - \bar N_n \right) 
+ \frac{1}{2} \, \sum\limits_{\ell = e, \mu, \tau} 
\left(N_{\nu_\ell} - \bar N_{\nu_\ell} \right) 
\right] , 
\label{eq:EnuLgen}
\end{eqnarray}
where the functions 
$N_e, N_p, N_n, N_{\nu_\ell}$ are the number densities of background electrons, 
protons, neutrons, and neutrinos, and $\bar N_e, \bar N_p, \bar N_n, 
\bar N_{\nu_\ell}$ are the densities of the corresponding antiparticles. 
To find the additional energy for antineutrinos, one should change the 
total sign in the right-hand side of Eq.~(\ref{eq:EnuLgen}). 

For a typical astrophysical medium, except for the early Universe and 
a supernova core, 
one has $\bar N_e \simeq \bar N_p \simeq \bar N_n \simeq N_{\nu_\ell}
\simeq \bar N_{\nu_\ell} \simeq 0$,
and $N_p \simeq N_e = Y_e\, N_B, \, N_n \simeq  (1-Y_e)\, N_B$, where $N_B$ 
is the barion density. One obtains
\begin{eqnarray}
W_e &=& \frac{G_{\rm F} \, N_B}{\sqrt{2}} 
\left(3\, Y_e-1 \right) ,
\label{eq:EnuLe}\\
W_{\mu,\tau} &=& - \frac{G_{\rm F} \, N_B}{\sqrt{2}} 
\left(1-Y_e \right) .
\label{eq:EnuLmu}
\end{eqnarray}
As $Y_e < 1$, the additional energy acquired by muon and tau left-handed 
neutrinos 
is always negative. At the same time, the additional energy of electron 
left-handed neutrinos becomes positive at $Y_e > 1/3$. 
And vice versa, the additional energy for electron antineutrinos is positive 
at $Y_e < 1/3$, while it is always positive for the muon and tauon 
antineutrinos. 
On the other hand, right-handed neutrinos and their antiparticles, left-handed 
antineutrinos, being sterile with respect to weak interactions, 
do not acquire an additional energy.

The additional energy $W$ from Eq.~(\ref{eq:EnuLe}) 
gives an effective mass squared $m_L^2$ to the left-handed neutrino, 
\begin{equation}
m_L^2 = {\cal P}^2 = (E + W)^2 - {\bf p}^2 \,, 
\label{eq:mL}
\end{equation}
where ${\cal P}$ is the neutrino four-momentum in medium, while 
$(E,\, {\bf p})$ would form the neutrino four-momentum in vacuum, 
$E = \sqrt{{\bf p}^2 + m_\nu^2}$. 

Given a $\nu \nu \gamma$ interaction, 
the additional energy of left-handed neutrinos in medium opens new 
kinematical possibilities for the radiative neutrino transition: 
\begin{equation}
\nu \to \nu + \gamma \,.
\label{eq:nunugamma}
\end{equation}

It should be self-evident, that the influence of the substance on the photon 
dispersion must be taken into account, 
$\omega = |{\bf k}|/n$, where $n \ne 1$ is the refractive index. 

First, a possibility exists that the medium provides the condition $n > 1$ 
(the effective photon mass squared is negative, $m_\gamma^2 \equiv 
q^2 < 0$) which corresponds to the well-known 
effect~\cite{Grimus:1993,D'Olivo:1996,Ioannisian:1997} of 
``neutrino Cherenkov radiation''.  
In this situation, the neutrino dispersion change under the medium influence 
is being usually neglected, because the neutrino dispersion is defined by the 
weak interaction while the photon dispersion is defined by the electromagnetic 
interaction. 

Pure theoretically, one more possibility could be considered when the photon 
dispersion was absent, and the process of the radiative neutrino transition
$\nu \to \nu \gamma$ would be caused by the neutrino dispersion only. 
As the left-handed neutrino dispersion is only changed, transitions become 
possible caused by the $\nu \nu \gamma$ interaction with the neutrino 
chirality change, e.g. due to the neutrino magnetic dipole moment. 

Just this situation called the ``spin light of neutrino'' ($SL \nu$), 
was first proposed and investigated in detail in an extended series of 
papers~\cite{Lobanov:2003,Lobanov:2004,Studenikin:2005,Lobanov:2005,Grigoriev:2005}.  
However, in the analysis of this effect the authors overlooked such an 
important phenomenon as plasma influence on the photon dispersion. 
As will be shown below, this phenomenon closes the $SL \nu$ effect for all 
real astrophysical situations. 

In this Letter, we reanalyse the process 
$\nu_L \to \nu_R \gamma$ taking into account both the neutrino 
dispersion and the photon dispersion in medium. 
Having in mind possible astrophysical applications, it is worthwhile to consider 
the astrophysical plasma as a medium, which transforms the photon into 
the plasmon, see e.g. Ref.~\cite{Braaten:1993} and the papers cited 
therein. 

To perform a kinematical analysis, it is necessary to evaluate the scales 
of the values of the left-handed neutrino additional energy $W$ and of the 
photon (plasmon) effective mass squared $m_\gamma^2$.

One readily obtains from Eq.~(\ref{eq:EnuLe}): 
\begin{equation}
W \simeq 6 \; {\rm eV}
\left(\frac{N_B}{10^{38} \, {\rm cm}^{-3}}\right) \left(3\, Y_e - 1 \right) ,
\label{eq:W}
\end{equation}
where the scale of the barion number density is taken, which is typical 
e.g. for the interior of a neutron star. 

On the other hand, a plasmon acquires in medium 
an effective mass $m_\gamma$ which is approximately constant at high energies. 
For the transversal plasmon, the value $m_\gamma^2$ is always positive, and 
is defined by the so-called plasmon frequency. 
In the non-relativistic classical plasma (i.e. for the solar interior) one has:
\begin{equation}
m_\gamma \equiv \omega_{\rm pl} = \sqrt{\frac{4 \pi \, \alpha \,N_e}{m_e}} \simeq 
4 \times 10^{2} \,{\rm eV}
\left(\frac{N_e}{10^{26} {\rm cm}^{-3}}\right)^{1/2}.
\label{eq:omega_pl_nr}
\end{equation}
For the ultra-relativistic dense matter one has:
\begin{equation}
m_\gamma =  \sqrt{\frac{3}{2}} \; \omega_{\rm pl} 
= \left(\frac{2 \, \alpha}{\pi} \right)^{1/2} 
\left(3\, \pi^2 \, N_e \right)^{1/3} \simeq 
10^{7} \,{\rm eV}
\left(\frac{N_e}{10^{37} \, {\rm cm}^{-3}}\right)^{1/3}.
\label{eq:omega_pl_r}
\end{equation}
In the case of hot plasma, when its temperature is the largest 
physical parameter, the plasmon mass is:
\begin{equation}
m_\gamma =  \sqrt{\frac{2 \, \pi \, \alpha}{3}} \; T
\simeq 
1.2 \times 10^{7} \,{\rm eV}
\left(\frac{T}{100 \, {\rm MeV}}\right).
\label{eq:omega_pl_h}
\end{equation}

One more physical parameter, a great attention was payed to in the $SL \nu$ 
analysis~\cite{Lobanov:2003,Lobanov:2004,Studenikin:2005,Lobanov:2005,Grigoriev:2005},
was the neutrino vacuum mass $m_\nu$. 
As the scale of neutrino vacuum mass could not exceed essentially 
a few electron-volts, which 
is much less than typical plasmon mass scales for real astrophysical situations, 
see Eqs.~(\ref{eq:omega_pl_nr})-(\ref{eq:omega_pl_h}), it is reasonable 
to neglect $m_\nu$ in our analysis.

Thus, in accordance with~(\ref{eq:mL}), a simple condition for the 
kinematic opening of the process $\nu_L \to \nu_R \gamma$ is:
\begin{equation}
m_L^2 \simeq 2 \, E \, W > m_\gamma^2 \,. 
\label{eq:mL>}
\end{equation}
This means that the process becomes kinematically opened when 
the neutrino energy exceeds the threshold value, 
\begin{equation}
E > E_0 = \frac{m_\gamma^2}{2 \, W} \,. 
\label{eq:p_0}
\end{equation}
Let us evaluate these threshold neutrino energies for different astrophysical 
situations. 

For the solar interior $N_B \simeq 0.9 \times 10^{26} \, {\rm cm}^{-3}$, 
$Y_e \simeq 0.6$, and the threshold neutrino energy is
\begin{equation}
E_0 \simeq 10^{10} \,{\rm MeV} \,, 
\label{eq:p_0_sun}
\end{equation}
to be compared with the upper bound $\sim$ 20 MeV for the solar neutrino energies. 

For the interior of a neutron star, where $Y_e \ll 1$, 
the Wolfenstein energy for neutrinos~(\ref{eq:EnuLe}), (\ref{eq:EnuLmu}) 
is negative, and the process $\nu_L \to \nu_R \gamma$ is closed. 
On the other hand, there exists a possibility for opening the antineutrino 
decay. Taking for the estimation $Y_e \simeq 0.1$, one obtains from~(\ref{eq:W}) 
and~(\ref{eq:omega_pl_r}) the threshold value
\begin{equation}
E_0 \simeq 10^{7} \,{\rm MeV} \,, 
\label{eq:p_0_NS}
\end{equation}
to be compared with the typical energy $\sim$ MeV of neutrinos emitted 
via the URCA processes. 

For the conditions of a supernova core, the additional energy of left-handed 
electron neutrinos can be obtained from Eq.~(\ref{eq:EnuLgen}) as follows:
\begin{equation}
W_e = \frac{G_{\rm F} \, N_B}{\sqrt{2}} 
\left(3\, Y_e + Y_{\nu_e} - 1 \right) ,
\label{eq:EnuLeSN}
\end{equation}
where $Y_{\nu_e}$ describes the fraction of trapped electron neutrinos 
in the core, $N_{\nu_e} = Y_{\nu_e}\, N_B$. 
Taking typical parameters of a supernova core, we obtain
\begin{equation}
E_0 \simeq 10^{7} \,{\rm MeV} \,, 
\label{eq:p_0_SN}
\end{equation}
to be compared with the averaged energy $\sim 10^{2}$ MeV of trapped neutrinos. 

In the early Universe, when plasma was almost charge symmetric, 
the Wolfenstein formula~(\ref{eq:EnuLgen}) giving zero should be changed 
to a more accurate expression for the additional energy which is identical 
for both neutrinos and antineutrinos~\cite{Notzold:1988,Elmfors:1996} 
\begin{equation}
W_i = - \frac{7 \, \sqrt{2} \, \pi^2 \, G_{\rm F} \, T^4}{45} 
\left( \frac{1}{m_Z^2} + \frac{2 \, \delta_{ie}}{m_W^2} \right) E \, .
\label{eq:W_early}
\end{equation}
The minus sign unambiguously shows that in the early Universe, in contrast 
to the neutron star interior, the decay process is forbidden 
both for neutrinos and antineutrinos. 

Thus, the above analysis shows that the nice effect of the 
``neutrino spin light'', unfortunately, has no place in real 
astrophysical situations because of the photon dispersion. 
The sole possibility for the discussed process $\nu_L \to \nu_R \gamma$ 
to have any significance could be connected only with the situation 
when an ultra-high energy neutrino threads a star. 
Obviously it could have only a methodical meaning. 
Let us calculate the process width for these purposes correctly. 

A neutrino having a magnetic moment $\mu_\nu$ interacts with photons, and the 
Lagrangian of this interaction is
\begin{equation}
{\cal L} = - \frac{i \, \mu_\nu}{2} \left( \bar \nu \sigma_{\alpha \beta} 
\nu \right) F^{\alpha \beta} \,,
\label{eq:L}
\end{equation}
where $\sigma_{\alpha \beta} = (1/2)\, (\gamma_\alpha \gamma_\beta - 
\gamma_\beta \gamma_\alpha)$, and 
$F^{\alpha \beta}$ is the tensor of the photon electromagnetic field. 

With the Lagrangian~(\ref{eq:L}), the invariant amplitude squared 
for the process $\nu_L \to \nu_R \gamma$, summarized over the transversal 
plasmon polarizations, can be obtained by the standard way:
\begin{equation}
|{\cal M}|^2 = 4 \, \mu_\nu^2 \, E^2 \left[ 2 \, W^2 \left(1 - \frac{\omega}{E} 
\right) - m_\gamma^2 \, \sin^2 \theta \right],
\label{eq:M^2}
\end{equation}
where $\omega$ is the plasmon energy, $\theta$ is the angle between the 
initial neutrino momentum ${\bf p}$ and the plasmon momentum ${\bf k}$. 
It should be stressed that discussing ultra-high energy neutrinos, and 
consequently the high plasmon energies, one can consider with a good 
accuracy the plasmon mass $m_\gamma$ as a constant depending on the 
plasma properties only, see Eqs.~(\ref{eq:omega_pl_nr})-(\ref{eq:omega_pl_h}). 
This is in contrast to the left-handed neutrino effective mass squared 
$m_L^2$, which is the dynamical parameter, see Eq.~(\ref{eq:mL}).  

The differential width of the process 
$\nu_L \to \nu_R \gamma$ is defined as:
\begin{equation}
{\mathrm{d}} \Gamma = \frac{|{\cal M}|^2}{8\, E \, (2 \pi)^2} 
 \, \delta (E + W - E' - \omega) \,
\delta^{(3)} ({\bf p} - {\bf p}' - {\bf k}) \, 
\frac{{\mathrm{d}}^3 p'\, {\mathrm{d}}^3 k}{E'\, \omega} \,,
\label{eq:dGamma}
\end{equation}
where the plasmon energy $\omega$ cannot be taken the vacuum one 
($\omega = |{\bf k}|$), as it was done in the $SL \nu$ 
analysis~\cite{Lobanov:2003,Lobanov:2004,Studenikin:2005,Lobanov:2005,Grigoriev:2005}, 
but it is defined by the dispersion 
in plasma, $\omega = \sqrt{{\bf k}^2 + m_\gamma^2}$. 

\begin{figure}[htb]
\centering
\includegraphics[width=0.6\textwidth]{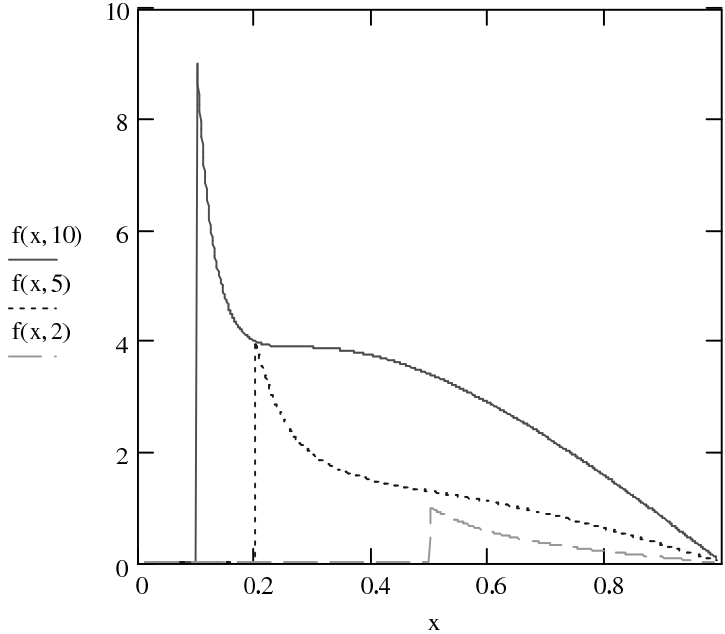}
\caption{The function $f (x, \varepsilon)$
defining the spectrum of plasmons from the 
left-handed neutrino decay for different values of the ratio 
$\varepsilon = E/E_0$ of the neutrino energy to the 
threshold neutrino energy, 
$\varepsilon = 10$ (solid line), 
$\varepsilon = 5$ (dotted line), 
and $\varepsilon = 2$ (dashed line).}
\label{fig:spectrum}
\end{figure}

Performing a partial integration in Eq.~(\ref{eq:dGamma}), one obtains for 
the photon (i.e. transversal plasmon) spectrum  
\begin{eqnarray}
{\mathrm{d}} \Gamma &=& \frac{\alpha}{4} 
\left(\frac{\mu_\nu}{\mu_{\rm B}}\right)^2 
\frac{m_\gamma^2 \, W}{m_e^2} 
\, f (x, \varepsilon) \, 
{\mathrm{d}} x  \qquad \left(\frac{1}{\varepsilon} \leqslant x \leqslant 1 \right) \,,
\nonumber\\
f (x, \varepsilon) &=& \varepsilon \, (1 - x) + 2 \left(1 
- \frac{\varepsilon + 1}{\varepsilon \, x} + 
\frac{1}{\varepsilon \, x^2}\right),
\label{eq:dGamma2}
\end{eqnarray}
where $\mu_{\rm B} = e/(2 \, m_e)$ is the Bohr magneton, and 
the notations are used $x = \omega/E$, and 
$\varepsilon = E/E_0$. Recall that $E_0 = m_\gamma^2/(2 \, W)$ is 
the threshold neutrino energy for the process to be opened. 
In Fig.~\ref{fig:spectrum} the function $f (x, \varepsilon)$ 
is shown for some values of the ratio $\varepsilon$.

Instead of the photon energy spectrum~(\ref{eq:dGamma2}), one can 
obtain also the spatial distribution of final photons. As the 
analysis shows, all the photons are created inside the narrow cone 
with the opening angle $\theta_0$, 
\begin{equation}
\theta < \theta_0 \simeq \frac{\varepsilon - 1}{\varepsilon} \, 
\frac{W}{m_\gamma}\,.
\label{eq:theta}
\end{equation}

Performing the final integration in Eq.~(\ref{eq:dGamma2}), one obtains 
the total width of the process
\begin{eqnarray}
\Gamma &=& \frac{\alpha}{8} 
\left(\frac{\mu_\nu}{\mu_{\rm B}}\right)^2 
\frac{m_\gamma^2 \, W}{m_e^2} \; F (\varepsilon) 
\qquad \left(\varepsilon \geqslant 1 \right)\,,
\nonumber\\
F (\varepsilon) &=& \frac{1}{\varepsilon} \left[(\varepsilon - 1) 
(\varepsilon + 7) - 4 (\varepsilon + 1)\, \ln \varepsilon \right] .
\label{eq:Gamma}
\end{eqnarray}

It should be noted that in the situation when $W < 0$, and the transition 
$\nu_L \to \nu_R \gamma$ is forbidden, the crossed channel 
$\nu_R \to \nu_L \gamma$ becomes kinemalically opened. As the analysis 
shows, the plasmon spectrum and the total decay width are described in 
this case by the same Eqs.~(\ref{eq:dGamma2}) and~(\ref{eq:Gamma}), 
with the only substitution $W \to |W|$. 

To illustrate the extreme weakness of the effect considered, let us 
evaluate numerically the mean free path of an ultra-high energy 
neutrino with respect to the radiative decay, when the neutrino threads 
a neutron star. For the typical neutron star parameters, 
$N_B \simeq 10^{38} \, {\rm cm}^{-3}$, $Y_e \simeq 0.05$, 
we obtain
\begin{equation}
L \simeq 10^{19} \, {\rm cm} \times 
\left(\frac{10^{-12}\,\mu_{\rm B}}{\mu_\nu}\right)^2 
\left[F \left(\frac{E}{10 \, {\rm TeV}} 
\right) \right]^{-1} \,, 
\label{eq:path}
\end{equation}
where the neutrino energy $E > E_0$, $E_0 \simeq 10$ TeV is the threshold 
energy for such conditions. 
The mean free path~(\ref{eq:path}) should be compared with the neutron 
star radius $\sim 10^{6}$ cm. 

In conclusion, we have shown that the effect of the 
``neutrino spin light'' has no place in real astrophysical situations 
because of the photon dispersion in plasma. 
The photon (plasmon) effective mass causes the threshold 
which leaves no room for the process. For a pure theoretical situation 
when an ultra-high energy neutrino threads a star, the total width of the 
process $\nu_L \to \nu_R \gamma$ is calculated with a correct taking 
account of the photon dispersion in plasma.
The extreme weakness of the effect considered is illustrated. 

\section*{Acknowledgements}

The work was supported in part 
by the Russian Foundation for Basic Research under the Grant No.~04-02-16253, 
and by the Council on Grants by the President of Russian Federation 
for the Support of Young Russian Scientists and Leading Scientific Schools of 
Russian Federation under the Grant No.~NSh-6376.2006.2.


\end{document}